\documentclass[graphicx, 12pt]{article}
\usepackage{indentfirst}

\usepackage{graphicx}
\usepackage{caption}
\usepackage{subfigure}

\begin{document}

\small
\hoffset=-1truecm
\voffset=-2truecm
\title{\bf The gamma-ray burst arising from neutrino pair annihilation
in the static and spherically symmetric black-hole-like wormholes}

\author{Yuxuan Shi, Hongbo Cheng\footnote
{E-mail address: hbcheng@ecust.edu.cn}\\
Department of Physics,\\ East China University of Science and
Technology,\\ Shanghai 200237, China\\
The Shanghai Key Laboratory of Astrophysics,\\Shanghai 200234,
China}

\date{}
\maketitle

\begin{abstract}
We look into the neutrino-antineutrino pair
($\nu+\bar{\nu}\longrightarrow e^{-}+e^{+}$) annihilation in the
Damour-Solodukhin wormhole spacetime whose metric component
involves a shift in contrast to the similar black hole. The deep
analysis of the surface temperature of the accretion disk of
static, spherically symmetric black-hole-like wormholes from R.
Kh. Karimov et.al. reveals that the accretion disks of the
wormholes are hotter than that of comparable black holes,
indicating that the wormholes accretion disk can release
neutrinos. Further we investigate the energy deposition rate from
the neutrino pair annihilation around the Damour-Solodukhin
wormhole thought as a mimicker of Schwarzschild black hole. By
comparison made between the black-hole-like wormhole and the
similar black hole, we demonstrate that the wormhole's accretion
disk drawing the annihilation can become a source of gamma-ray
burst although the more significant deviation from the similar
black hole reduces the emitted power slightly. The ratio of energy
deposition per unit time from the annihilation surrounding the
accretion disk of the Damour-Solodukhin wormhole over the emitting
power of black hole might alter noticeably depending on how
slightly the metrics of the wormhole differ from the black hole
spacetime.
\end{abstract}

\vspace{-0.5cm} \hspace{0cm} PACS number(s): 

\noindent \textbf{I.\hspace{0.4cm}Introduction}

The black holes are solutions to the Einstein equations
theoretically and are also accepted as compact objects in
astrophysics [1-3]. The detections of the candidates of black
holes such as supermassive ones and Kerr's ones proceed [4, 5].
The traversable wormhole with nontrivial topological structures
can alternatively be known as the solutions to the Einstein
equation [6, 7]. It should be pointed out that the weak energy
condition must be violated at the throat of any static and
spherically symmetric traversable wormhole within the frame of
general relativity [7]. Under the weak energy condition the
authors of Ref.[8] found the wormhole solutions in the alternative
gravity. Based on the field equations in the general relativity or
alternative gravity, the solutions like black holes or wormholes
were obtained [1, 7]. It was found that the black holes and
wormholes are characterized by an event horizon and throat
respectively according to the phenomena observed and the
theoretical influence [1-6]. The properties of wormholes need to
be explored in various directions. In order to wonder how the
electromagnetic rays pass through the source throat in the strong
gravitational field, more efforts have been contributed to the
wormholes, including deflection angles [9], visualizations [10,
11], gravitational lensing [12-15], wave optics [16], etc.. A lot
of explorations have been contributed to the orbits of stars or
pulsars around the black holes at the centre of galaxy to detect
the relevant wormholes [17]. Among the wormholes, a kind of models
named as black hole foils were considered and the so-called
black-hole-like wormholes have no event horizon but a throat [18].
The fact that the globally static wormholes resemble the similar
black holes is noteworthy [18]. Some physicists focused on this
kind of wormholes on purpose to distinguish them from the black
holes that are similar to them. The authors of Ref.[6, 18]
discussed the Hawking's radiation from the wormholes to reveal
that the radiation is too weak to be measured. It was found that a
tidal force acting on a falling body near the Damour-Solodukhin
wormhole is greater than that near the Schwarzschild black hole
[19]. The Kerr-type wormholes' accretion disks and their images
were examined [20, 21]. The quasinormal modes for the
Damour-Solodukhin wormholes were derived and calculated [22, 23].
The analysis is performed on gravitational lensing and the shadow
image in the wormholes [24, 29]. The gravitational-wave echoes by
the wormhole attract more attentions [30-33]. The collision of two
test particles was scrutinized around the Damour-Solodukhin
wormhole which is similar to the Schwarzschild black hole and it
was exhibited that the centre-of-mass energy for the head-on
collision of the particles is large so long as the tiny deviation
of wormhole metric from black hole spacetime exist [34].

In order to understand the gamma-ray burst, a volume of researches
have been conducted on its adequate source [35, 36]. The course of
neutrino-antineutrino annihilation converting the
electron-positron pairs thanks to the hot accretion disk around
the gravitational source. In the background of black hole, the
accretion disk particles move inwards in the nearly geodesic
circular orbits within the equatorial plane [35, 36]. During the
spiral motion, the radiant heat due to stress and dynamical
friction will be emitted from the disk surface [35, 36]. For a
thin accretion disk surrounding a black hole or the other compact
objects, the time averaged energy flux, the disk temperature, the
differential luminosity, the conversion efficiency of accreting
mass into radiation should be elaborated [35, 36]. The hot
accretion disks of black holes produce neutrino-antineutrino pairs
and the pair annihilations may lead to the gamma-ray bursts [37,
38]. The sufficiently hot thin accretion disk with accretion rate
as $\dot{M}~(0.1-1)M_{\odot s^{-1}}$ around the black hole can
emit neutrinos and antineutrinos, where $M_{\odot}$ is mass of the
sun [37, 38]. The neutrino pairs can annihilate above the disk  to
convert into electron-positron pairs [39, 42]. It is necessary for
the bursts proceeding as powerful cosmic explosions that the
energy deposition rate arising from electron-positron must be
large enough [39-42]. The gamma-ray bursts may be thought as
systems powered by newborn stellar-mass black holes accreting
matter at hyper-critical rates [39, 43-55]. The annihilation
process mentioned above brings about a greater advance in the
neutrino heating of the envelope resulting in a supernova
explosion, further the pair $e^{-}e^{+}$ near the surface of
collapsing neutron star, the output of the process, releases the
powerful gamma rays that may provide a possible explanation of the
observed bursts [56]. It was found that the efficiency of
neutrino-antineutrino annihilation into electron-positron pair
increases over the Newtonian values up to a factor of 4 or 30 in
the case of Type II supernovae and close to the surface of
collapsing neutron stars respectively [56]. The researches on the
neutrino pair annihilation near the black holes also continue. It
was found that the off-axis contribution to the energy-momentum
deposition rate from the $\nu\bar{\nu}$ pair collision above a
Kerr black hole within a thin accretion disk is larger by a factor
of 10~20 than the on-axis energy deposition rate [37, 38, 40, 57].
In addition, the generalization of the general relativity cause
the energy deposition processes involving the
neutrino-antineutrino annihilation into electron-positron pairs in
the vicinity of accretion disks surrounding the compact objects to
increase energy flux significantly [40, 58]. The neutrino pair
annihilation efficiency in the background of the astronomical
objects swallowing an $f(R)$ global monopole is much greater and
this kind of stellar bodies can be well qualified as a source of
heavier gamma-ray bursts [59]. However few studies have reported
on the annihilation in the environment of wormholes, the
black-hole-like wormholes in particular.

It is significant to investigate the neutrino-antineutrino
annihilation near the wormholes. We could examine the process
around the black-hole-like wormholes to compare our results such
as energy deposition rate with those from the black holes. Our
findings may be useful in clarifying their difference. The aim of
this paper is to examine the reaction
$\nu+\bar{\nu}\longrightarrow e^{-}+e^{+}$ in the spherically
symmetric Schwarzschild-type wormholes. Based on the fact that the
sufficiently hot accretion disk emits neutrinos and antineutrinos
[43-46], we aim at the disk of the black-hole-like wormhole. We
wonder whether the accretion disk belonging to this kind of
wormholes can generate the neutrinos. Secondly, we derive the
integral form of the neutrino pair annihilation efficiency subject
to the Damour-Solodukhin wormholes. Further we compute the ratio
of total energy deposition over the Newtonian ones for factor
$\Lambda$ as a correction to the Schwarzschild spacetime. Our
numerical estimation will relate to the factor $\Lambda$. The
results depicted in the figures will indicate the possibility that
this kind of Damour-Solodukhin wormhole possessing the
annihilation can serve as sources of gamma-ray burst while may
help to reveal how different the black holes and wormholes are
from each other in terms of the emitted power from neutrino
reaction. The discussion will proceed and the conclusions will be
listed in the end.

\vspace{0.8cm} \noindent \textbf{II.\hspace{0.4cm}The accretion
disk around the Damour-Solodukhin wormholes}

As mentioned above that only sufficiently hot accretion disk of
compact body can set free the neutrinos [35-38], we are going to
follow the procedure of Ref. [20] to elaborate the surface
temperature of the disk of Damour-Solodukhin wormholes. We start
to report on the geodesics for a particle motion in the background
of Damour-Solodukhin wormhole [18, 20]. The line element of the
spacetime is given by [18],

\begin{eqnarray}
ds^{2}=g_{\mu\nu}dx^{\mu}dx^{\nu}\hspace{5cm}\nonumber\\
=(f(r)+\Lambda^{2})dt^{2}-\frac{dr^{2}}{f(r)}-r^{2}(d\theta^{2}
+\sin^{2}\theta d\varphi^{2})
\end{eqnarray}

\noindent where $f(r)=1-\frac{2GM}{r}$ with Newton constant $G$
and a positive mass parameter $M$. The small and positive
dimensionless parameter $\Lambda$ causes the metric to be
different from the Schwarzschild's one, without or with event
horizon respectively [18]. This kind of metric (1) has the throat
at $r=2GM$ that joins two isometric, asymptotically flat regions
[18]. The radial coordinate $r$ is restricted in a range $2GM\leq
r<\infty$ [18]. According to the spherically symmetric spacetime
like Eq.(1), we can limit the particle motion to the equatorial
plane as $\theta=\frac{\pi}{2}$ while $\frac{d\theta}{d\tau}=0$.
The Lagrangian as the description of the planar motion is given by
[18, 60],

\begin{equation}
\mathcal{L}=\frac{1}{2}(f(r)+\Lambda^{2})(\frac{dt}{d\tau})^{2}
-\frac{1}{2}\frac{1}{f(r)}(\frac{dr}{d\tau})^{2}-\frac{1}{2}r^{2}
(\frac{d\varphi}{d\tau})^{2}
\end{equation}

\noindent where $\tau$ is the proper time. The 4-momentum of the
particle is
$p_{\mu}=\frac{\partial\mathcal{L}}{\partial\frac{dx^{\mu}}{d\tau}}$
and the components are,

\begin{equation}
p^{t}=(f(r)+\Lambda^{2})\frac{dt}{d\tau}=E
\end{equation}

\begin{equation}
p^{r}=\frac{1}{f(r)}\frac{dr}{d\tau}\hspace{2cm}
\end{equation}

\begin{equation}
p^{\varphi}=r^{2}\frac{d\varphi}{d\tau}=L\hspace{1.5cm}
\end{equation}

\noindent The momenta satisfy $p_{\mu}p^{\mu}=\eta$ [60]. The
geodesics introduce $E=constant$ and $L=constant$ [60]. The
particle moves along the planar geodesics in the equatorial plane
governed by the metric (1) and is described by the equation of
radial motion with momenta (3)-(5) as follow [60],

\begin{equation}
(\frac{dr}{d\tau})^{2}+V_{eff}(r)=0
\end{equation}

\noindent where the effective potential is,

\begin{equation}
V_{eff}(r)=f(r)[\eta-\frac{E^{2}}{f(r)+\Lambda^{2}}+\frac{L^{2}}{r^{2}}]
\end{equation}

\noindent In view of the metric function $f(r)$, the effective
potential can be formulated as [18],

\begin{equation}
V_{eff}(r)=(1-\frac{2GM}{r})(\eta-\frac{E^{2}}{1+\Lambda^{2}
-\frac{2GM}{r}}+\frac{L^{2}}{r^{2}})
\end{equation}

\noindent which is consist with results without the spin of
wormhole [20]. In the case of geodesics which the particles move
along, the parameter $\eta=0$ or $\eta=1$ corresponds to massless
or massive particles respectively. The effective potential (8) can
be shown in Figure 1. A series of curves of the potential are
similar. It should be pointed out that the zeros of the effective
potential exist and the values of the zeros decrease with
increasing $\Lambda$.

According to the techniques of Ref.[60], we can combine the
momentum components in Eq.(3) and Eq.(5). We calculate the angular
velocity [20],

\begin{eqnarray}
\Omega=\frac{\frac{d\varphi}{d\tau}}{\frac{dt}{d\tau}}\hspace{2cm}\nonumber\\
=\frac{L}{E}\frac{f(r)+\Lambda^{2}}{r^{2}}\hspace{0.7cm}\nonumber\\
=\frac{L}{E}\frac{1+\Lambda^{2}-\frac{2GM}{r}}{r^{2}}
\end{eqnarray}

Now we focuss on the radial motion equation (6). If the particles
move circularly, the derivative $\frac{dr}{d\tau}$ will vanish,
leading $V_{eff}(r)=0$ [60]. From Figure 1, the effective
potentials have minima and the magnitudes of minima depend on the
variable $\Lambda$, the deviation of wormholes from black holes.
The effective potential requires that the circular orbits exist at
its local minimum like $\frac{dV_{eff}(r)}{dr}=0$ [60]. The radius
of the innermost stable circular geodesic orbit (ISCO) on the
accretion disk satisfies
$\frac{d^{2}V_{eff}(r)}{dr^{2}}|_{r=r_{ISCO}}=0$ [61]. We can
gather [60, 61],

\begin{eqnarray}
\{V_{eff}(r)=0\\
\frac{dV_{eff}(r)}{dr}=0\\
\frac{d^{2}V_{eff}()r}{dr^{2}}=0
\end{eqnarray}

\noindent We make derivation on the effective potential two times
and plot the derivative $\frac{d^{2}V_{eff}()r}{dr^{2}}$ in the
Figure 2. The values of zeros become smaller with larger
$\Lambda$. The Figure 2 reports that the function
$\frac{d^{2}V_{eff}(r)}{dr^{2}}$ tends to vanish when the radial
coordinate approaches to the infinity. We substitute the effective
potential (8) into the conditions (10)-(12) for the ISCO [20],

\begin{eqnarray}
V_{eff}(r)=0\hspace{2.5cm}\nonumber\\
=f(r)[1-\frac{E^{2}}{f(r)+\Lambda^{2}}+\frac{L^{2}}{r^{2}}]
\end{eqnarray}

\begin{eqnarray}
\frac{dV_{eff}(r)}{dr}=0\hspace{2cm}\nonumber\\
=\frac{f'(r)}{(f(r)+\Lambda^{2})^{2}}E^{2}-\frac{2}{r^{3}}L^{2}
\end{eqnarray}

\begin{eqnarray}
\frac{d^{2}V_{eff}(r)}{dr^{2}}=0\hspace{4.5cm}\nonumber\\
=[\frac{f''(r)}{(f(r)+\Lambda^{2})^{2}}
-\frac{2(f'(r))^{2}}{(f(r)+\Lambda^{2})^{3}}]E^{2}
+\frac{6}{r^{4}}L^{2}
\end{eqnarray}

\noindent According to the Eq.(11) and Eq.(12), we find that [20],

\begin{equation}
E^{2}=\frac{(1+\Lambda^{2}-\frac{2GM}{r})^{2}}
{1+\Lambda^{2}-\frac{3GM}{r}}
\end{equation}

\begin{equation}
L^{2}=\frac{GMr}{1+\Lambda^{2}-\frac{3GM}{r}}
\end{equation}

\noindent We make use of the Eq.(13) and metric function
$f(r)=1-\frac{2GM}{r}$ to obtain the equation which the radius of
ISCO obeys [20],

\begin{equation}
2E^{2}M^{2}(1+\Lambda^{2})r^{4}-3L^{2}(1+\Lambda^{2})^{3}r^{3}
+18L^{2}(1+\Lambda^{2})^{2}Mr^{2}-36L^{2}(1+\Lambda^{2})M^{2}r
+24L^{2}M^{3}=0
\end{equation}

\noindent The Eq.(18) relates the radius of ISCO to the parameters
including the characteristic variable $\Lambda$. The relation is
depicted in Figure 3. The so-called ISCO radius is a decreasing
function of the deviation from the similar black hole, the smaller
ISCO radius under the larger $\Lambda$.

Under the laws of energy and angular momentum conservation, the
derivative is [36, 62],

\begin{equation}
\frac{dL_{\infty}}{d\ln r}=4\pi r\sqrt{-g}kF(r)
\end{equation}

\noindent Here $L_{\infty}$ is the luminosity at infinity. The
function of radial coordinate $F(r)$ is the flux of radiant energy
emitted from the upper surface of the accretion disk. The flux of
radiant energy can be expressed in the following integral form
with the specific energy $E$ and the specific angular momentum $L$
from Eq.(14) and Eq.(15) respectively [36, 62],

\begin{equation}
F(r)=-\frac{\dot{M}}{4\pi\sqrt{-g}}\frac{1}{(E-\Omega h)^{2}}
\frac{d\Omega}{dr}\int_{r_{ISCO}}^{r}(E-\Omega L)\frac{dL}{dr}dr
\end{equation}

\noindent The dot means the derivative with respect to the
time.

The accretion disk can be thought as a model of black body
radiation and the surface temperature is given by [36],

\begin{equation}
T(r)=(\frac{F(r)}{\sigma})^{\frac{1}{4}}
\end{equation}

\noindent where $\sigma$ is Stefan-Boltzmann constant. We have to
draw the surface temperature again in the Figure 4 which is the
same as the part of Figure 2 in Ref. [20]. Our repetition helps us
to explain the impacts of the surface temperature profiles. The
temperatures are high enough to encourage the accretion of matter
down a hole to proceed irresistibly. The curves of accretion disk
temperature rises as the dimensionless parameter $\Lambda$ becomes
larger. The Figure 4 as well as the Figure 2 from Ref. [20]
displays that the wormhole disk is about $4.4\%$ hotter than it is
in the Schwarzschild black hole's immediate vicinity. It could be
emphasized that the wormholes will continue releasing the neutrino
pairs. It is even more possible for the wormholes to keep emission
of the neutrino pairs whose annihilation can become the source of
gamma ray burst.

\vspace{0.8cm} \noindent \textbf{III.\hspace{0.4cm}The energy
deposition rate by the neutrino annihilation process in the
Damour-Solodukhin spacetime}

In view of the discussion on accretion disk [18, 20] in the last
section, we compute the energy deposition rate by the
neutrino-antineutrino pair annihilation around the static,
spherically symmetric black-hole-like wormholes to wonder whether
this kind of disks attracting the annihilation may become sources
of gamma-ray burst. We pay attention to the particle motion with a
circular orbit in the background of Damour-Solodukhin wormhole
with the description of metric (1). During the process of emission
of neutrinos from the wormhole-inspired accretion disks, the
energy deposition rate per unit time and unit volume is written as
[38],

\begin{equation}
\frac{dE}{dtdV}=\frac{21\pi^{4}\zeta(5)}{h^{6}}KG_{F}^{2}
F(r)(kT)^{9}
\end{equation}

\noindent where $h$ is the Planck constant and $k$ is the
Boltzmann constant. $G_{F}$ is the Fermi constant. Here $K$
represents $K(\nu_{\mu}, \bar{\nu}_{\mu})$, $K(\nu_{\tau},
\bar{\nu}_{\tau})$, $K(\nu_{e}, \bar{\nu}_{e})$ respectively and
is shown in the expressions [56, 63],

\begin{eqnarray}
K(\nu_{\mu}, \bar{\nu}_{\mu})=K(\nu_{\tau},
\bar{\nu}_{\tau})\hspace{1.5cm}\nonumber\\
=\frac{1}{6\pi}(1-4\sin^{2}\theta_{W}+8\sin^{4}\theta_{W})
\end{eqnarray}

\noindent and

\begin{equation}
K(\nu_{e},
\bar{\nu}_{e})=\frac{1}{6\pi}(1+4\sin^{2}\theta_{W}+8\sin^{4}\theta_{W})
\end{equation}

\noindent with the Weinberg angle $\sin^{2}\theta_{W}=0.23$. The
angular integration factor is denoted as
$F(r)=\frac{2\pi^{2}}{3}(1-x)^{4}(x^{2}+4x+5)$ [56]. Here
$x=\sin^{2}\theta_{r}$, where $\theta_{r}$ represents the angle
between the trajectory and the tangent velocity [66] and can be
defined as
$\tan\theta_{r}=\frac{v_{r}}{v_{\varphi}}=\frac{\sqrt{g_{11}(r)}}{r}
\frac{dr}{d\varphi}$ with radial component $v_{r}$ and angular
ones $v_{\varphi}$ of velocity respectively and further
$\frac{1}{r}\frac{dr}{d\varphi}=\frac{\tan\theta_{r}}{\sqrt{g_{11}(r)}}$
[58]. We can hire the null geodesic equation
$(\frac{1}{r^{2}}\frac{dr}{d\varphi})=\frac{1}{b^{2}\frac{1}{g_{11}^{2}(r)}}
-\frac{1}{r^{2}}\frac{1}{g_{11}(r)}$ [1, 60] to reformulate the
angle $\theta_{r}$ like
$\frac{r\cos\theta_{r}}{\sqrt{g_{00}(r)}}=b$ with the impact
parameter $b$ while
$b=\frac{r_{0}\cos\theta_{r_{0}}}{\sqrt{g_{00}(r_{0})}}$. Here the
combination $\frac{r\cos\theta_{r}}{\sqrt{g_{00}(r)}}=
\frac{r_{0}\cos\theta_{r_{0}}}{\sqrt{g_{00}(r_{0})}}$ with
$\theta_{r_{0}}=0$ because of the photosphere tangent to the
stellar surface at radius $r_{0}$ leading,

\begin{equation}
\cos
\theta_{r}=\frac{r_{0}}{r}\sqrt{\frac{g_{00}(r)}{g_{00}(r_{0})}}
\end{equation}

\noindent where $r_{0}$ is the neutrinosphere radius. According to
the metric function (1),
$x^{2}=1-\frac{r_{0}^{2}}{r^{2}}\frac{f(r)+\Lambda^{2}}{f(r_{0})+\Lambda^{2}}$
From Eq.(20) the local temperature $T=T(r)$ measured by a observer
satisfies $T(r)\sqrt{g_{00}(r)}=constant$ [56]. Here $g_{00}(r)$
is a component of spacetime metric like Eq.(1). We can perform the
integration of the rate per unit time and volume formulated in
Eq.(20) on the spherically symmetric volume around the wormhole to
declare the total amount of energy converted from neutrinos to
electron-positron pairs [64],

\begin{eqnarray}
\dot{Q}_{w}=\frac{dE}{\sqrt{g_{00}}dt}\hspace{5cm}\nonumber\\
=\frac{56\pi^{7}\zeta(5)}{h^{6}}KG_{F}^{2}k^{9}(\frac{7}{4}\pi
ac)^{-\frac{9}{4}}L_{\infty}^{\frac{9}{4}}r_{0}^{-\frac{3}{2}}\hspace{1cm}\nonumber\\
\times(g_{00}(r_{0}))^{\frac{9}{4}}\int_{1}^{\infty}(x-1)^{4}
(x^{2}+4x+5)\nonumber\\
\times\frac{y^{2}\sqrt{-g_{11}(r_{0}y)}}{(g_{00}(r_{0}y))^{\frac{9}{2}}}dy
\end{eqnarray}

\noindent where $y=\frac{r}{r_{0}}$. The term $\frac{7}{4}\pi ac$
is from the luminosity for a single neutrino species at the
neutrinosphere [56]. In order to illustrate the influence from the
factor $\Lambda$ on the energy conversion per unit time, we can
compare the energy deposition rate (24) with ones in Newton's
gravity [58],

\begin{eqnarray}
\frac{\dot{Q}_{W}}{\dot{Q}_{Newt}}\hspace{5.5cm}\nonumber\\
=3(g_{00}(r_{0}))^{\frac{9}{4}}\int_{1}^{\infty}(x-1)^{4}
(x^{2}+4x+5)\nonumber\\
\times\frac{y^{2}\sqrt{-g_{11}(r_{0}y)}}{(g_{00}(r_{0}y))^{\frac{9}{2}}}
dy
\end{eqnarray}

We can also proceed with the angular integration to obtain the
derivative $\frac{d\dot{Q_{W}}}{dr}$ as a function of the radial
coordinate $r$ to exhibit the enhancement according to the scheme
of Ref.[58],

\begin{eqnarray}
\frac{d\dot{Q_{W}}}{dr}\hspace{8.5cm}\nonumber\\
=\frac{56\pi^{7}\zeta(5)}{h^{6}}KG_{F}^{2}k^{9}(\frac{7}{4}\pi
ac)^{-\frac{9}{4}}L_{\infty}^{\frac{9}{4}}(x-1)^{4}(x^{2}+4x+5)
r_{0}^{-\frac{5}{2}}\nonumber\\
\times(\frac{g_{00}(r_{0})}{g_{00}(r)})^{\frac{9}{4}}\sqrt{-g_{11}(r)}
(\frac{r}{r_{0}})^{2}\hspace{4cm}
\end{eqnarray}

We quantify the ratio $\frac{\dot{Q}_{W}}{\dot{Q}_{Newt}}$ in the
vicinity of gravitational source described with metric (1) and
plot the dependence of the ratio on $\frac{r_{0}}{M}$ in the
Figure 5. In contrast to the Newton case, it is obvious that the
energy deposition rates have been augmented owing to the
wormholes, especially within the region $\frac{r_{0}}{M}<4$. We
find that the emission powers of neutrino pair annihilation refer
to the wormholes are distinct from the powers in the cases of
black holes. It should be pointed out that the larger variable
$\Lambda$ gives rise to smaller ratio
$\frac{\dot{Q}_{W}}{\dot{Q}_{Newt}}$. The Figure 6 also states
that the more considerable deviations from black holes hinder the
efficiency of transmitting energy for the black-hole-like
wormholes specified by Eq.(1). According to the numerical
estimation from Ref.[65], the parameter $\Lambda$ has to be
selected extremely small, even exponentially small in the case of
evaporation. So long as the modification $\Lambda\neq 0$, the
gravitational sources have throats instead of even horizon even if
the value of $\Lambda$ is tiny. If the parameter $\Lambda$
modifying the spacetime structure of the black hole is
sufficiently small, it is difficult to distinguish a wormhole from
a black hole astrophysically through the measurement of Hawking's
radiation [18]. We compare the Damour-Solodukhin wormhole with
Schwarzschild black hole in the Figure 7 which declares that the
difference between the magnitudes of energy deposition per unit
time from the annihilation for Schwarzschild black holes and black
hole-typed wormholes respectively is manifest within the region of
factor $\Lambda$, which can help us to distinguish the wormholes
from the similar black holes. The Figure 5 also displays that the
ratio of emitted powers is large enough when $\Lambda<0.3$ and the
black-hole-like wormholes accreting the neutrino annihilation can
trigger the gamma-ray burst. The derivative
$\frac{d\bar{Q}_{W}}{dr}$ as a function of radial coordinate is
shown in the Figure 8 and the amplification is greatest close to
the neutron star's surface.

Here we need to confirm whether this kind of wormholes holding the
neutrino pairs annihilation can act as a source of gamma-ray
burst, so we concentrate on the efficiency of the energy released
from the annihilation. The amount of emitted power must be huge
enough. The works on the emitted spectrum are also significant. We
should mention that the dependence of the energy per unit time
from the process on the frequency will be considered in future.

\vspace{0.8cm} \noindent \textbf{IV.\hspace{0.4cm}Discussion and
Conclusion}

We study the neutrino pair annihilation
$\nu\bar{\nu}\longrightarrow e^{-}e^{+}$ around the
Damour-Solodukhin wormhole. According to Ref. [20], we analyze the
surface temperature of the accretion disk of the gravitational
source with throat to find that the factor $\Lambda$ inserted into
the Schwarzschild metric change the temperature as Ref.[18]. The
wormholes with larger $\Lambda$ have hotter accretion disks which
can emit more neutrinos [18]. We derive and calculate the emitted
power $\dot{Q}_{W}$ from the neutrino-antineutrino pair
annihilation in the background of $\Lambda$-corrected wormhole to
find that this kind of wormholes can lead a huge increase in the
conversion efficiency of energy from the disk although the factor
$\Lambda$ alleviates the energy deposition rate. We also show that
the accretion disk of the so-called black-hole-like wormhole
attracting the neutrino annihilation can be proposed as a
well-qualified candidate of gamma-ray burst. In addition, it
should be emphasized that the difference between the magnitudes of
energy deposition per unit time from the annihilation for
Schwarzschild black hole and black-hole-typed wormhole
respectively is manifest within the region of factor $\Lambda$,
which can help us to distinguish the wormholes from black holes.

\vspace{1cm}
\noindent \textbf{Acknowledge}

This work is partly supported by the Shanghai Key Laboratory of
Astrophysics 18DZ2271600.

\newpage
\begin{figure}
\setlength{\belowcaptionskip}{10pt} \centering
\includegraphics[width=15cm]{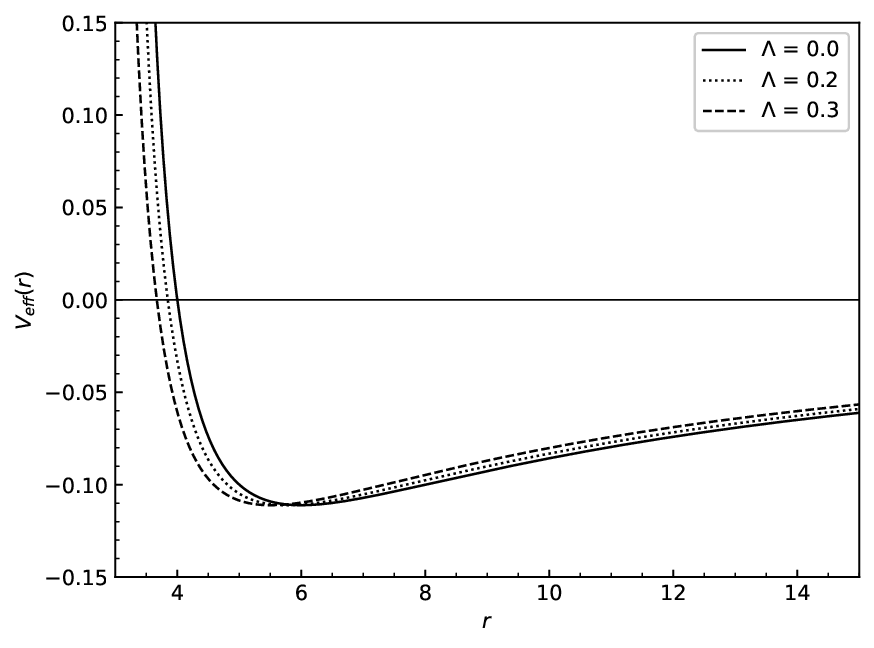}
\caption{The effective potential as function of radial coordinate
for $\Lambda=0, 0.1, 0.3$ shown with solid, dotted and dashed
curves respectively.}
\end{figure}

\newpage
\begin{figure}
\setlength{\belowcaptionskip}{10pt} \centering
\includegraphics[width=15cm]{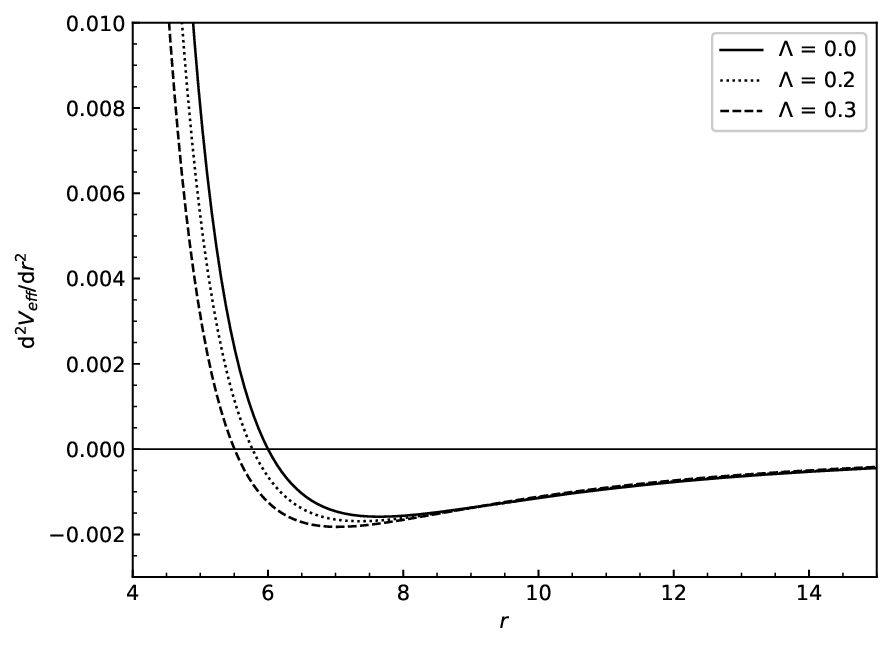}
\caption{The derivatives for effective potential
$\frac{d^{2}V_{eff}(r)}{dr^{2}}$ for $\Lambda=0, 0.1, 0.3$ given
by the solid, dotted and dashed curves respectively.}
\end{figure}

\newpage
\begin{figure}
\setlength{\belowcaptionskip}{10pt} \centering
\includegraphics[width=15cm]{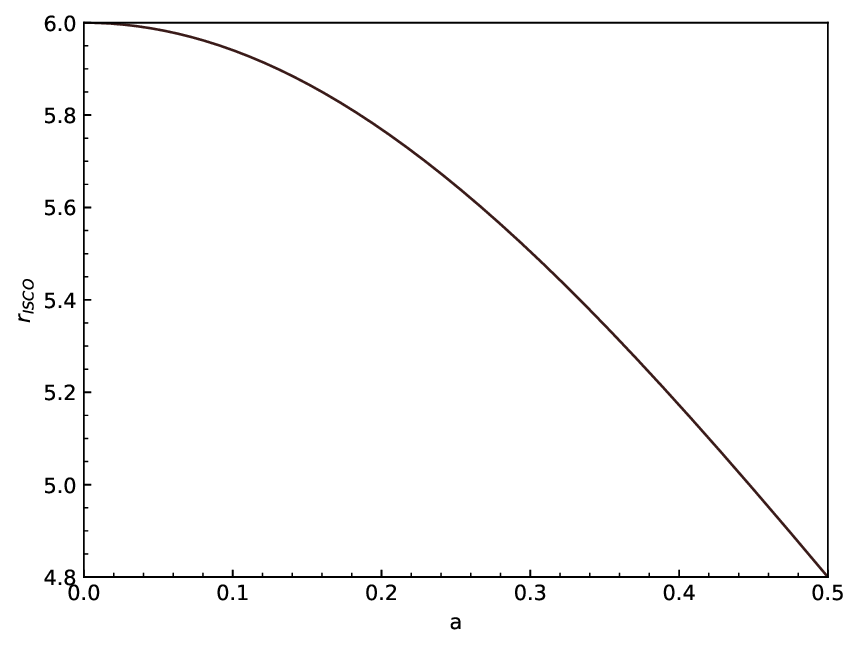}
\caption{The profile of the ISCO radius versus the parameter
$\Lambda$ as deviation from the similar black hole.}
\end{figure}

\newpage
\begin{figure}
\setlength{\belowcaptionskip}{10pt} \centering
\includegraphics[width=15cm]{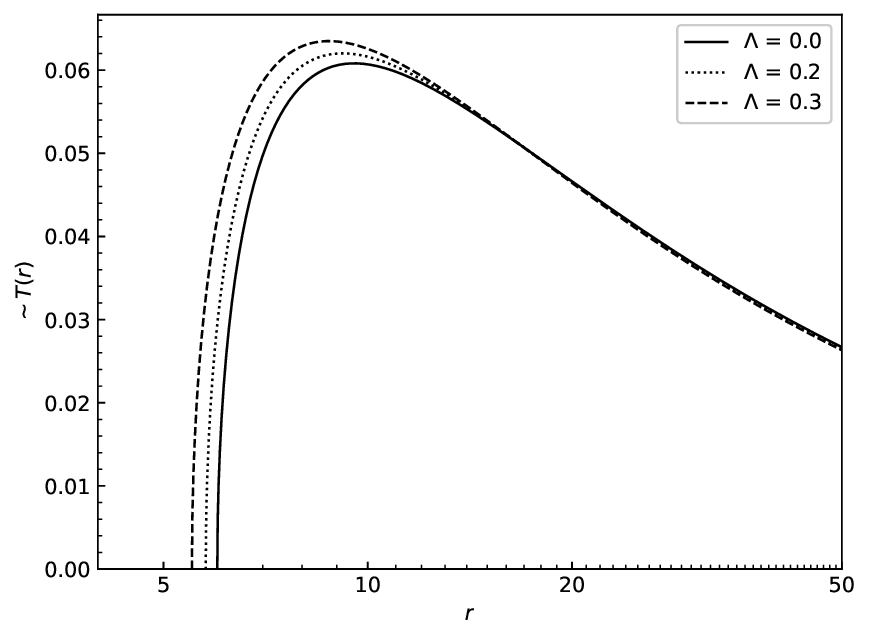}
\caption{The solid, dotted, dashed curves of the temperature
depending on the distance away from a thin accretion disk around a
black-hole-typed wormhole for $\Lambda=0, 0.1, 0.3$ respectively.}
\end{figure}

\newpage
\begin{figure}
\setlength{\belowcaptionskip}{10pt} \centering
\includegraphics[width=15cm]{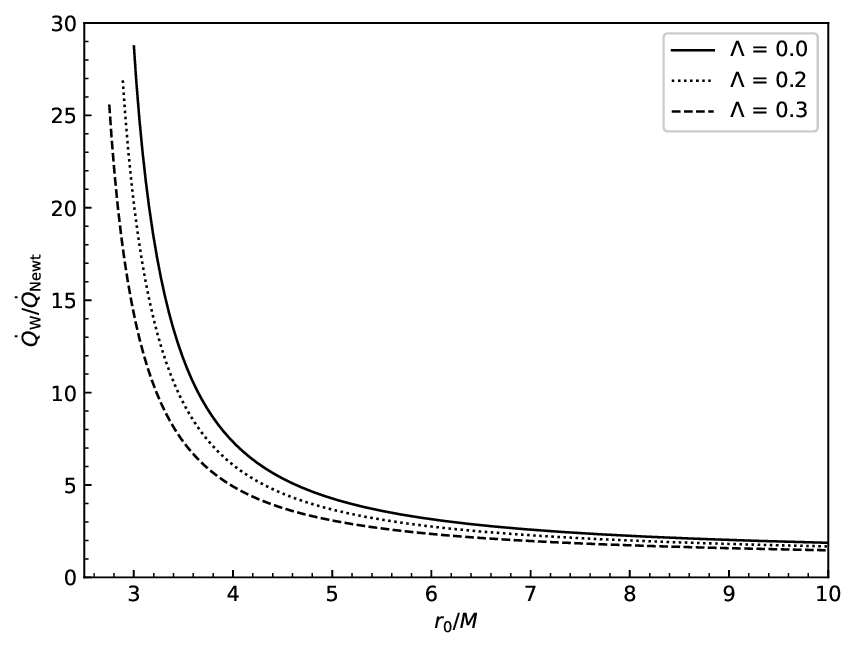}
\caption{The solid, dotted and dashed curves of the ratio
$\frac{\dot{Q_{W}}}{\dot{Q_{Newt}}}$ as functions of the ratio
$\frac{r_{0}}{M}$ for factors $\Lambda=0, 0.1, 0.3$ respectively.}
\end{figure}

\newpage
\begin{figure}
\setlength{\belowcaptionskip}{10pt} \centering
\includegraphics[width=15cm]{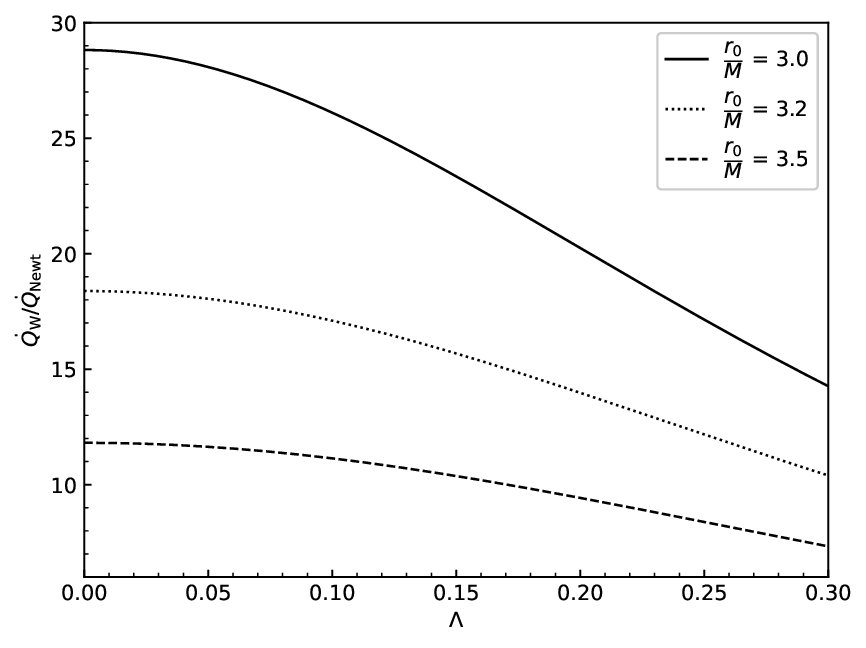}
\caption{The ratio $\frac{\dot{Q_{W}}}{\dot{Q_{Newt}}}$ for
factors $\Lambda$ depicted by the solid, dotted and dashed curves
at $\frac{r_{0}}{M}=3, 3.2, 3.5$ respectively.}
\end{figure}

\newpage
\begin{figure}
\setlength{\belowcaptionskip}{10pt} \centering
\includegraphics[width=15cm]{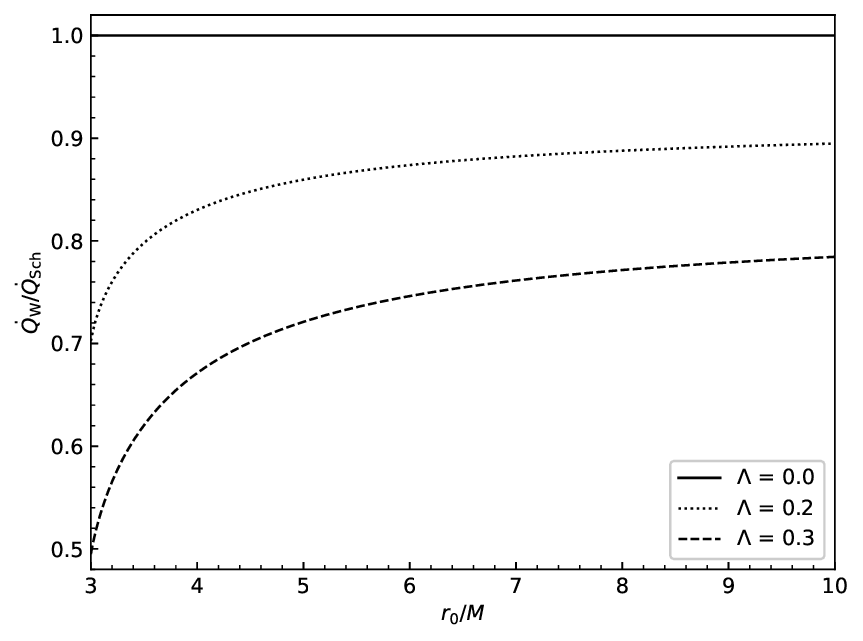}
\caption{The solid, dotted and dashed curves of ratio
$\frac{\dot{Q_{W}}}{\dot{Q_{Sch}}}$ for factors $\Lambda=0, 0.2,
0.3$ respectively.}
\end{figure}

\newpage
\begin{figure}
\setlength{\belowcaptionskip}{10pt} \centering
\includegraphics[width=15cm]{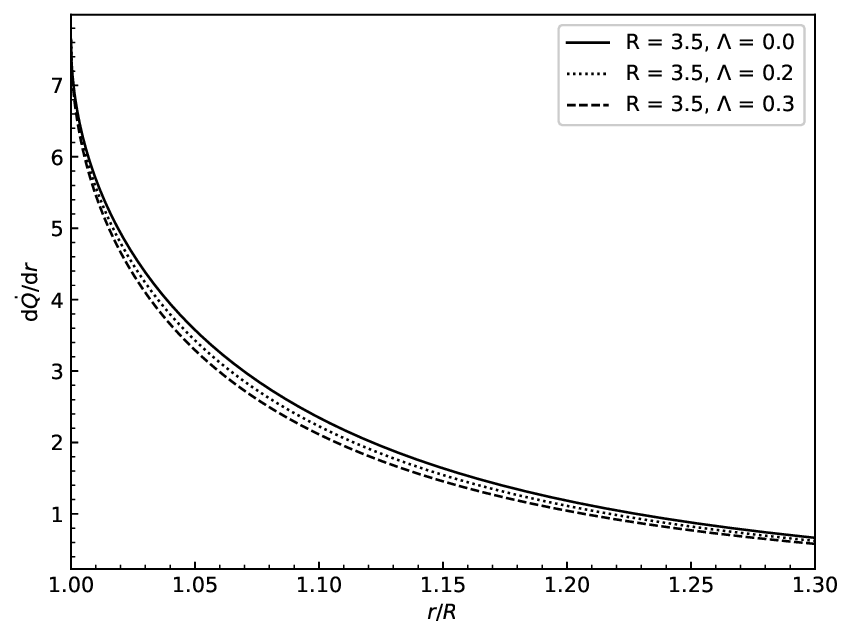}
\caption{The solid, dotted, dashed curves of the derivative
$\mathrm{d}\dot{Q}_{W}/\mathrm{d}r$ as a function of the radius of
the wormhole under the factors $\Lambda=0, 0.2, 0.3$
respectively.}
\end{figure}

\end{document}